# Highlights

**Finding regions of bounded motion in binary asteroid environment using Lagrangian descriptors**

S. Raffa, G. Merisio, F. Topputo

- Bounded motion in non-autonomous dynamics is sought with Lagrangian descriptors.
- Lagrangian descriptors are compared against well-established variational methods.
- The Didymos binary system exhibits regions of bounded motion about Dimorphos.
- Solar radiation pressure breaks most of the Didymos system dynamical structures.
- Lagrangian descriptors are computationally inexpensive dynamical indicators.

# Finding regions of bounded motion in binary asteroid environment using Lagrangian descriptors

Sebastiano Raffa[a,*], Gianmario Merisio[a,**] and Francesco Topputo[a]

[a]*Department of Aerospace Science and Technology, Politecnico di Milano, Via La Masa 34, Milano 20156, Italy*

ABSTRACT

Trajectory design in highly-perturbed environments like binary asteroids is challenging. It typically requires using realistic, non-autonomous dynamical models in which periodic solutions derived in autonomous systems vanish. In this work, Lagrangian descriptors are employed in the perturbed planar bi-elliptic restricted four-body problem to find regions of bounded motion over a finite horizon about Dimorphos, the secondary body of the (65803) Didymos binary system. Results show that Lagrangian descriptors successfully reveal phase space organizing structures both in the unperturbed and perturbed planar bi-elliptic restricted four-body problem. With no solar radiation pressure, regions of bounded motion are visually identified, so granting access to a vast selection of bounded orbits about Dimorphos. Conversely, the presence of solar radiation pressure breaks down the majority of structures, leading to a large region of unstable motion with rare exceptions. Lagrangian descriptors are computationally inexpensive dynamical indicators that could be conveniently applied to astrodynamics.

## 1. Introduction

A binary asteroid environment is characterized by a low-gravity field where small dynamical perturbations affect significantly the dynamics [10, 11]. Consequently, trajectory design for space exploration missions targeting small solar system bodies is challenging, especially for spacecraft with a limited control authority. In the near future, two missions part of the Asteroid Impact and Deflection Assessment (AIDA) international collaboration [4] are scheduled to visit the (65803) Didymos binary system: NASA's DART [5], and ESA's Hera [33]. The former is expected to conduct a kinetic impact experiment on the smaller body Dimorphos, while the latter is expected to study the effects of the impact. They both plan on deploying CubeSats in the proximity of the Didymos system, therefore raising the need for methods to find quasi-periodic, bounded orbits in the proximity of the double asteroid.

The planar circular restricted three-body problem (PCR3BP) is far from being representative of the dynamical environment close to the Didymos–Dimorphos system. A perturbed planar bi-elliptic restricted four-body problem (PBER4BP) including solar radiation pressure (SRP) is better suited for a more accurate representation of the real dynamics. However, periodic orbits derived in the PCR3BP around Didymos vanish in the PBER4BP since the problem is non-autonomous. They are replaced by quasi-periodic solutions.

In this work, regions of bounded motion for a spacecraft flying in the proximity of the Didymos–Dimorphos binary system are computed through a purely numerical approach. We tackle the problem computing a dynamical indicator on large grids of initial conditions (ICs). In previous works, the method was successfully applied with several indicators defined by: the Fourier analysis of the solutions, e. g., frequency map [26, 27]; the variation of phase-space variables during the motion, e. g., sup-map analysis [13] and Lagrangian descriptors (LDs) [31]; the solutions of variational equations, e. g., Lyapunov indicators [1], finite-time Lyapunov exponent (FTLE) [15, 41], mean exponential growth factor of nearby orbits (MEGNO) [6], and fast Lyapunov indicator (FLI) [14]. Several comparison papers are available in the literature [8, 19, 28].

The choice of an indicator is goal oriented. To illustrate, frequency map discerns regions of regular and chaotic motion [26, 27]. Similarly, Lyapunov indicators discriminate between regular and chaotic motions. Additionally, they can successfully compute stable, unstable, and Lagrangian manifolds [1]. Sup-map analysis [13] and LDs [31] rely on tailored integral norms computed along the motion. FLI informs on the precision loss during the numerical integrations [14, 17]. Modified FLIs identify chaotic regions and $L_1$–$L_2$ manifolds when applied to the three-body problem [18].

The goal of the paper is to reveal regions of bounded motion in double asteroid environment using LDs [29, 31]. For this purpose, distant retrograde orbits (DROs) about the secondary and Lyapunov orbits (LOs) of Lagrange point $L_1$ are first computed in the PCR3BP through differential correction [23, 40] and inspected against the LD scalar field for classification. The LD scalar field of the unperturbed and perturbed PBER4BP is visually examined to identify solutions that persist about Dimorphos, regardless of the perturbations. Results indicate that regions of bounded motion still exist in the PBER4BP. However, they are almost completely lost in presence of the SRP. Overall, LDs provide insightful dynamical information and are computationally inexpensive, so being an alternative to other chaos indicators typically used in astrodynamics [15, 32, 42, 43].

The remainder of the paper is organized as follows. In Section 2, the dynamical model is described. The methodology is discussed in Section 3. Results are shown in Section 4. Eventually, conclusions are drawn in Section 5.

## 2. Equations of motion

The perturbed PBER4BP describes the motion of a particle in a gravitational field generated by three bodies moving in elliptic orbits. Let the primaries be Didymos (D1) and Dimorphos (D2). The model is expressed in the synodic reference frame centered at the primaries barycenter, which rotates and pulsates to keep their distance equal to one [24]. Let $\mu = m_{D2}/(m_{D1} + m_{D2})$, where $m_{D1}$ and $m_{D2}$ are the masses of D1 and D2, respectively. The positions of D1 and D2 are $(-\mu, 0)$ and $(1-\mu, 0)$, respectively. The equations of motion (EoM) are scaled such that the sum of D1 and D2 masses is set to one as well as their distance, and their period is scaled to $2\pi$ [24]. By designating the primaries true anomaly $f$ as the independent variable of the system, the EoM are [11, 24]

$$\ddot{\mathbf{r}} + \begin{bmatrix} 0 & -2 \\ 2 & 0 \end{bmatrix} \dot{\mathbf{r}} = \nabla\Omega - \alpha \left( \frac{\mathbf{r}_S}{\|\mathbf{r}_S\|^3} + \frac{\mathbf{r} - \mathbf{r}_S}{\|\mathbf{r} - \mathbf{r}_S\|^3} \right) + \beta \frac{\mathbf{r} - \mathbf{r}_S}{\|\mathbf{r} - \mathbf{r}_S\|^3} \tag{1}$$

$$\dot{\theta} = \gamma \frac{(1 + e_S \cos \theta)^2}{(1 - e_S^2)^{3/2}} \frac{(1 - e_D^2)^{3/2}}{(1 + e_D \cos f)^2} \tag{2}$$

where $(\dot{\cdot})$ and $(\ddot{\cdot})$ denote the first and second derivatives with respect to the true anomaly $f$; $\theta$ is the true anomaly of the double asteroid barycenter (D) with respect to the Sun (S) and its derivative is obtained through the chain rule [24]; $\mathbf{r} = (x, y)$ and $\mathbf{r}_S = (x_S, y_S)$ are the nondimensional position vectors of the spacecraft and the Sun, respectively, expressed in the synodic reference frame; $\mu_{(\cdot)}$, $a_{(\cdot)}$, and $e_{(\cdot)}$ refer to the gravitational parameter, the semi-major axis, and the eccentricity of the body $(\cdot)$, respectively (see Table 1). Then, $\Omega$ is the potential function that reads [24]

$$\Omega = \frac{1}{1 + e_D \cos f} \left[ \frac{1}{2}(x^2 + y^2) + \frac{1 - \mu}{r_{D1}} + \frac{\mu}{r_{D2}} + \frac{1}{2}\mu(1 - \mu) \right] \tag{3}$$

where $r_{D1}$ and $r_{D2}$ are the distances from Didymos and Dimorphos, respectively. The nondimensional coefficients $\alpha$, $\beta$, and $\gamma$ are computed as

$$\alpha = \frac{\text{TU}^2}{\text{LU}^3}\mu_S = \frac{\mu_S}{\mu_D(1 + e_D \cos f)}, \quad \beta = \frac{\text{TU}^2}{\text{LU}^3}\frac{P_0 d_{\text{AU}}^2 C_r A}{m} = \frac{P_0 d_{\text{AU}}^2 C_r A}{m\mu_D(1 + e_D \cos f)}, \quad \text{and } \gamma = \sqrt{\frac{\mu_S + \mu_D}{\mu_D}\left(\frac{a_D}{a_S}\right)^3} \tag{4}$$

where $P_0 = 4.56\,\text{N}\,\text{km}^{-2}$ is the solar radiation pressure at 1 AU [7]; $d_{\text{AU}} = 149\,597\,870.700\,\text{km}$ is the Astronomical Unit [7]; $C_r = 1.2$ is the assumed reflectivity coefficient; $A = 1.8\,\text{m}^2$ is the assumed Sun-projected area on the spacecraft for SRP evaluation; $m = 10\,\text{kg}$ is the assumed spacecraft mass. Consequently, the numerical values of the coefficients are $\alpha = [3.657\,723, 3.883\,974] \times 10^{18}$, $\beta = [6.075\,314, 6.451\,107] \times 10^{17}$, and $\gamma = 6.529\,288 \times 10^{-4}$. Coefficients $\alpha$ and $\beta$ are expressed as ranges because of their dependence through the true anomaly $f$. TU and LU are the normalization factors given by

$$\text{TU} = \sqrt{\frac{a_D^3}{\mu_D}\frac{(1 - e_D^2)^{3/2}}{(1 + e_D \cos f)^2}}, \quad \text{and} \quad \text{LU} = \frac{a_D(1 - e_D^2)}{1 + e_D \cos f}. \tag{5}$$

Finding regions of bounded motion in binary asteroid environment using Lagrangian descriptors

**Table 1**
Physical parameters.

| Parameter[1] | Unit | Value | Reference |
|---|---|---|---|
| $\mu$ | - | $9.214\,228 \times 10^{-3}$ | |
| $\mu_D$ | km$^3$ s$^{-2}$ | $3.522\,601 \times 10^{-8}$ | [11] |
| | LU$^3$ TU$^{-2}$ | $[9.708\,738 \times 10^{-1},\ 1.030\,928]$ | |
| $\mu_S$ | km$^3$ s$^{-2}$ | $1.327\,124 \times 10^{11}$ | |
| | LU$^3$ TU$^{-2}$ | $[3.657\,723,\ 3.883\,974] \times 10^{18}$ | |
| $a_D$ | km | 1.19 | |
| | LU | $[9.708\,738 \times 10^{-1},\ 1.030\,928]$ | |
| $a_S$ | km | $2.460\,287 \times 10^8$ | Horizons System[2] |
| | LU | $[2.007\,251,\ 2.131\,41] \times 10^8$ | |
| $e_D$ | - | 0.03 | |
| $e_S$ | - | 0.383\,638 | |

The Sun position vector is retrieved according to

$$\mathbf{r}_S = -\frac{1}{\text{LU}} \frac{a_S(1-e_S^2)}{(1+e_S\cos\theta)} \begin{bmatrix} \cos f & \sin f \\ -\sin f & \cos f \end{bmatrix} \begin{bmatrix} \cos\theta \\ -\sin\theta \end{bmatrix}. \quad (6)$$

The PCR3BP is recovered setting $\alpha = 0$, $\beta = 0$, and $e_D = 0$ in Eqs. (1), and (3), and by replacing the true anomaly $f$ with the nondimensional time $t$. The EoM have been integrated with a multistep, variable-step, variable-order, Adams–Bashforth–Moulton, predictor-corrector solver of orders 1st to 13th with both relative and absolute tolerances set to $10^{-12}$ [30].

## 3. Methodology
### 3.1. Lagrangian descriptors

LDs provide insight that appears to be linked with the geometric pattern of structures that govern transport in phase space. Their definition and heuristic arguments explaining why they are effective are presented in [31]. A theoretical framework is discussed in [29]. However, the connection between LDs and geometric patterns governing the transport in phase space is controversial and largely disputed in the literature. Indeed, LDs are not derived from mathematically well defined variational principles, thus their relation to invariant manifolds is unclear and mathematically not well defined [19, 22]. Moreover, LDs are not objective, i. e., structures resulting from the scalar field depend on the frame of the observer, whereas material curves such as periodic orbits are frame-indifferent [38, 39]. Finally, counter-examples to the method of Lagrangian descriptors are discussed in the literature. Specifically, they face smooth contour lines of LD at invariant manifolds, singular features of LD at irrelevant points, and failure when dealing with Hamiltonian systems [38]. In this study, we used the following LD definition

$$M(\mathbf{x}_0, f_f) = \int_0^{f_f} |\mathcal{F}(\mathbf{x}(f))|\, df \quad (7)$$

where $|\mathcal{F}(\mathbf{x}(f))| = \|\dot{\mathbf{x}}\|$ is a positive bounded scalar, where $\mathbf{x} = [x, y, \dot{x}, \dot{y}]^\top$ is the phase space state and $\dot{\mathbf{x}} = [\dot{x}, \dot{y}, \ddot{x}, \ddot{y}]^\top$ is its derivative. In [31], an extensive class of different LDs was defined based on the integrand, the selected norm, and the integration interval. In this work, we use one LD that is computed with forward integration, since we are interested in the future evolution of ICs. The LD is computed appending its integrand to the space state equations with a zero initial value and propagating the extended dynamics.

---

[1] The subscripts D and S represent quantities of the Didymos–Dimorphos and Didymos–Dimorphos barycenter about the Sun orbital motions, respectively.

[2] https://ssd.jpl.nasa.gov/horizons/ [last accessed March 1, 2022].

## 3.2. Distant retrograde orbits and Lyapunov orbits

DROs are periodic solutions of the PCR3BP in which the motion in the synodic frame about the secondary is retrograde, as opposed to the rotation of the secondary around the primary [2]. Their motion and stability have been studied in numerous works [3, 16, 35, 36]. In this study, DROs are used to classify the regions featured by the LD.

We compute DROs through a differential correction and numerical continuation technique, which exploits the symmetric features of the PCR3BP and proves effective in identifying periodic solutions [23, 40]. DROs are obtained starting from a mirror configuration of the form $\mathbf{x}_0 = [x_0, 0, 0, \dot{y}_0]^\top$ and stopping at the next $x$-axis crossing occurring at $t = T/2$ [37], $T$ being the orbital period. At the crossing, a second mirror configuration $\mathbf{x}_{T/2} = [x_{T/2}, 0, 0, \dot{y}_{T/2}]^\top$ is enforced maintaining $x_0$ while iteratively updating the velocity along the $y$-axis as follows

$$\dot{y}_0^{(k+1)} = \dot{y}_0^{(k)} - \left(\phi_{34}^{(k)}\right)^{-1} \dot{x}_{T/2}^{(k)} \tag{8}$$

where $\phi_{ij}$ is the element $(i, j)$ of the state transition matrix evaluated at $t = T/2$. The complete family is built through numerical continuation increasing $x_0$, and guessing $\dot{y}_0$ to be equal to the last IC corrected.

LOs are families of planar periodic orbits surrounding the collinear Lagrange points $L_1$, $L_2$, and $L_3$ [7]. In this study, LOs about the Lagrange point $L_1$ are obtained through differential correction following the same procedure previously introduced for the computation of DROs but using appropriate initial guesses.

## 3.3. Finding regions of bounded motion

Our study is focused on the qualitative motion of orbits, therefore the primaries are assumed to be point masses and physical impacts with them are not checked. The devised methodology, applied to the (65803) Didymos binary system case study, is made of two stages:

i) classification of the regions featured by the LD scalar field computed in a simplified, autonomous dynamics;

ii) identification of the bounded motion regions of interest exploiting the LD scalar field propagated in a more representative, non-autonomous dynamical model.

In this study, bounded motion regions are those islands of ICs that produce bounded orbits over the finite horizon considered when computing the LD.

Stage i) is carried out overlapping on the Poincaré section $y = \dot{x} = 0$ families of periodic DROs and $L_1$ LOs computed in the PCR3BP to the selected LD scalar field. This serves to understand the behavior of the discovered regions which is unknown a priori. The family of periodic orbits is expected to fall within a well-resolved region of the LD scalar field. In stage ii), the phase space is sampled relying on the previously acquired information. Dynamical models with an increasing degree of fidelity are progressively explored. Firstly, the unperturbed PBER4BP described by Eqs. (1), and (2) with $\beta = 0$ is investigated. Then, the PBER4BP perturbed by the SRP is examined. If regions of bounded motion still exist, they are expected to be similar to the ones found in step i). On the contrary, when they are lost, the LD scalar field is supposed to have a completely different look. LD scalar fields are computed in the $x\dot{y}$-plane on grids of $400 \times 400$ points. ICs are propagated for ten primaries revolutions, hence from $f_0 = 0$ to $f_f = 20\pi$. The selected finite horizon corresponds to approximately five days. Such time span is larger than typical durations (i. e., 3-4 days) of trajectory arcs considered in mission profiles for CubeSats flying in the proximity of a binary asteroid [10, 11].

## 4. Results

Results of stage i) are shown in Fig. 1. The red curve representing the DROs family lies within the smooth region that extends from the bottom-right to the top-left corners of the plot. The region narrows towards higher velocities and further distances from the secondary body. Many features characterized by large LD values surround the smooth blue area in which the red curve lies. Five trajectories are sampled on the Poincaré section, three from the smooth region, and two from the rippled areas (see ICs in Table 2). The first three (samples a, b, and e) behave similarly to the DROs family. Conversely, samples c and d diverge.

According to [31], LDs can effectively separate regions of different qualitative motion. Based solely on LDs, only separatrices (or boundaries) of the phase space over the finite horizon considered can be identified. They correspond to abrupt changes in a LD field, where in general the derivative of the LD transverse to the separatrix is discontinuous at the separatrix itself [29, 31]. An additional step is required to label identified regions as islands of either bounded or unbounded motion. A classification should be performed leveraging some previous knowledge of the problem. An approach could be studying trajectories sampled from across the separatrices or within the different identified areas and classify them according to a criterion. Trajectories are already available since they are required for the computation of the LD. Alternatively, the problem can be studied in a simplified model (e. g., an autonomous dynamics), so classifying the regions relying on insight obtained by other techniques (e. g., invariant manifolds, periodic orbit families). Both ways allow overcoming the impossibility to a priori tell which regions feature bounded or unbounded motion.

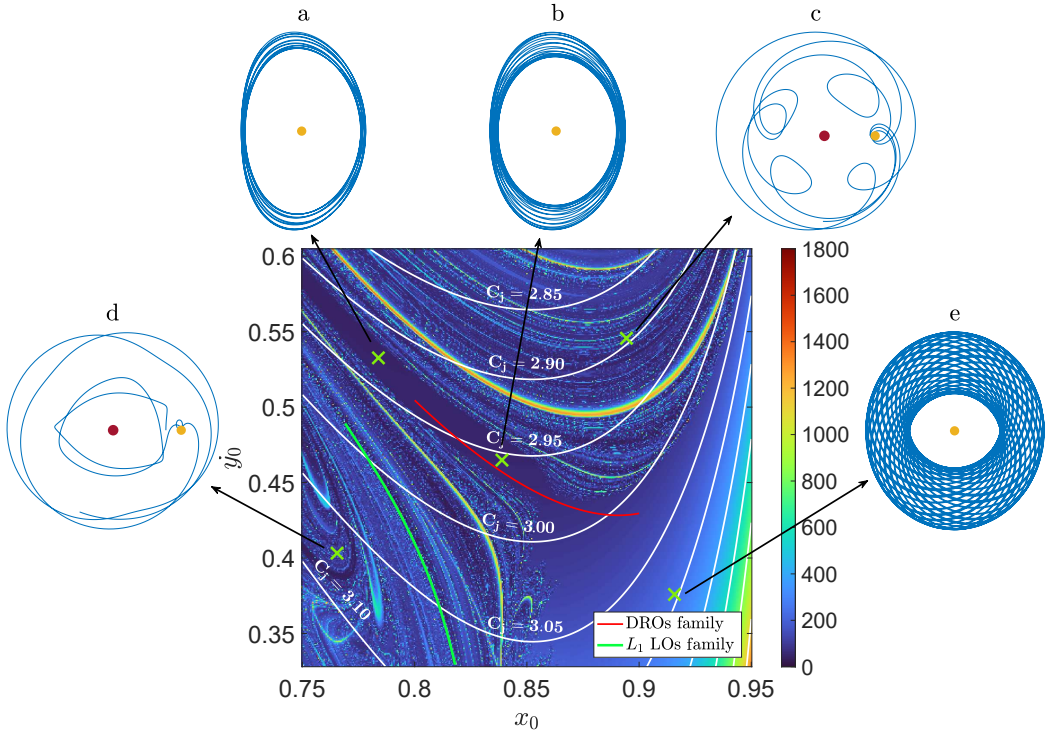

**Figure 1:** Periodic orbit families overlapped to LD scalar field in the PCR3BP, Poincaré section $y_0 = \dot{x}_0 = 0$. Level curves of Jacobi constant $C_j$ are plotted in white. The DROs family lies within a smooth region. Samples a, b, and e evolve in bounded orbits. Samples c and d escape. Didymos and Dimorphos represented not to scale as red and yellow dots, respectively.

In this work, the classification step carried out in the PCR3BP is performed specifically to acquire the necessary knowledge required to visually identify islands of bounded and unbounded motion when LD scalar fields are propagated in more representative models (non-autonomous dynamics), where techniques commonly applied in the PCR3BP fail. For instance, smooth areas appear to correspond to islands of bounded motion in the problem under study. Conversely, areas featuring many abrupt changes seem to correlate to unbounded motion. In Fig. 1, the smooth blue area contains ICs that persistently revolve about Dimorphos. It is clearly distinct from rippled zones where unbounded motion is observed. In practice, investigation over coarse grids of large domains could be performed for a fast identification of boundaries and potential regions of interest, and to extract fundamental information. After such preliminary analysis, more mathematically reliable tools (e. g., FTLEs, FLIs, variational theory for Lagrangian coherent structures) can be exploited for an unbiased identification of features in the phase space based on the specific needs of the application.

A panel showing the maximum distance from the asteroids barycenter in the considered ten primaries revolutions is presented in Fig. 2. The plot is of interest because the goal of the paper is to find islands of bounded motion. The patterns in Fig. 2 strongly resembles those in Fig. 1. The distances of four samples orbits (A, B, C and D) are reproduced as a function of the true anomaly $f$. Orbits A, C and D escape from Dimorphos, while orbit B remains bounded.

The presence of the $L_1$ LOs family visible in Fig. 1 offers the chance to test the LD method capability to detect stable invariant manifolds of the system. The $L_1$ LO orbit with Jacobi constant $C_j = 3.155\,086$ is selected as test case. Results are shown in Fig. 3. The stable manifold $\mathcal{W}^s_{L_1}$ and the unstable manifold $\mathcal{W}^s_{L_1}$ are plotted in Fig. 3a over the finite horizon $f \in [0, 8]$. The LD scalar field propagated over the same finite horizon is presented in Fig. 3b together with the stable

Finding regions of bounded motion in binary asteroid environment using Lagrangian descriptors manifold cut $\Gamma^s_{L_1}$ on the $x\dot{y}$-plane (see the bottom-left corner). The field is computed only for a grid of ICs having $y_0 = 0$ and $C_j = 3.155\,086$ (that of the selected $L_1$ LO orbit). The area where no real solutions of $\dot{x}_0 = (2U - C_j - \dot{y}_0)^{1/2}$ exist is labeled 'forbidden' and colored in gray [7]. Note that $U = (x^2 + y^2)/2 + (1-\mu)/r_{D1} + \mu/r_{D2} + \mu(1-\mu)/2$. Results show that structures visible in Fig. 3b correlate to the stable manifold cut $\Gamma^s_{L_1}$. The cut corresponds to the intersection of the stable manifold with the $x$-axis (see the purple segment in Fig. 3a). Furthermore, level curves of the gradient vector magnitudes of the LD field are shown in the top-left corner of Fig. 3 for an improved visualization of the abrupt changes in the field. The gradient vector field is calculated with central differences for interior points and single-sided differences for points along edges.

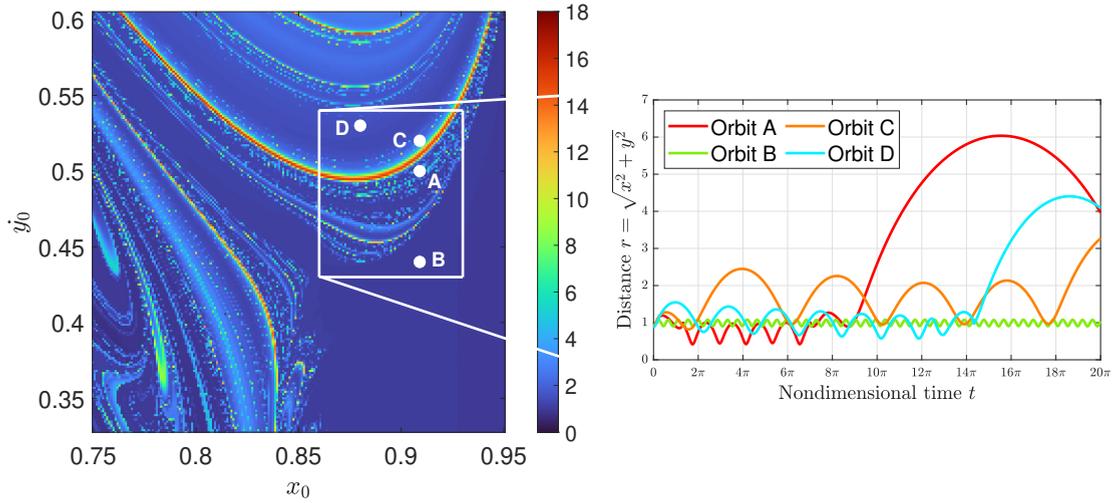

**Figure 2:** Maximum distance from asteroids barycenter over ten revolutions in the PCR3BP, Poincaré section $y_0 = \dot{x}_0 = 0$. Panel computed on a grid of $200 \times 200$ points. The distances from asteroids barycenter of four sample orbits are shown as a function of true anomaly $f$. Orbits A, C and D escape from Dimorphos, while orbit B remains bounded.

Results of stage i) are validated against a variational diagnostic. Specifically, results provided in Figs. 1 and 3b are replicated using the FTLE technique [15, 20, 41] instead of the LD method. Analysis are carried out over the same finite horizon, thus propagating ICs for ten revolutions of the primaries. A FTLE is defined as $\Lambda(\mathbf{x}_0, f_0; f_f) = \log \lambda_n(\mathbf{x}_0, f_0; f_f)/2/(f_f - f_0)$, where $\lambda_n$ is the maximum eigenvalue of the Cauchy–Green strain tensor [21]. Computing FTLEs requires propagating the variational equations of the PCR3BP [34], computing the Cauchy–Green strain tensor, and solving for the maximum eigenvalue [15, 41]. The outcome is presented in Fig. 4. Very similar patterns to those already seen in Figs. 1 and 2 are found even in Fig. 4a. Likewise, Fig. 4b shows how the FTLE technique detects structures in correspondence of features visible in Fig. 3b.

In stage ii), the fidelity of the non-autonomous dynamical model is progressively increased. The LD field propagated in the PBER4BP ($\beta = 0$) with the Sun initial true anomaly $\theta_0 = 0$ is reported in Fig. 5. Compared to Fig. 1, the wide smooth region still exists but new features emerge on the frontier with rippled areas. Furthermore, the channel containing the periodic DROs family is narrowed. Again, five samples are picked on the Poincaré section: three in the smooth region (samples f, g, and j), and two in the rippled areas located aside the central channel (samples h, and i). As before, the former exhibit bounded motion, while the latter immediately escape. Their ICs are reported in Table 2.

The LD field obtained moving to the PBER4BP perturbed by the SRP is presented in Fig. 6. Results with Sun initial true anomalies $\theta_0 = 0$, and $\pi$ are shown in Figs. 6a and 6b, respectively. In presence of the SRP, many new features emerge and the previously discovered smooth region characterized by bounded motion is completely lost. The tested ICs evolve into unstable orbits but sample o, which flies very close to Dimorphos. Remarkably, samples k and l are qualitatively really similar besides being located very far from each other. They belong to the same region, proving once again how effective LDs are in separating areas by different dynamical behaviors. As expected, almost all regions of bounded motion vanished when the Didymos system is at the pericenter ($\theta_0 = 0$) of its heliocentric orbit. Indeed, previous works already pointed out how the SRP sweeps away the spacecraft under the natural dynamics in

Finding regions of bounded motion in binary asteroid environment using Lagrangian descriptors

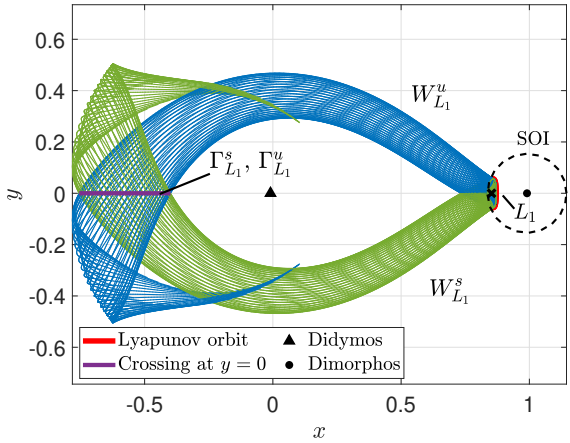
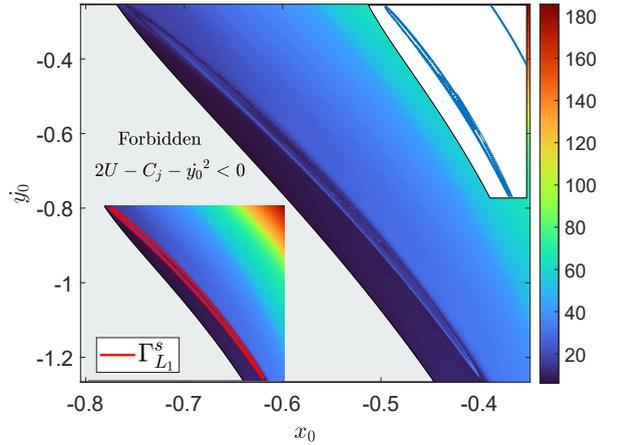

(a) Invariant manifolds associated to $L_1$ LO (in red). Stable (in green) and unstable (in blue) invariant manifolds propagated from $f_0 = 0$ to $t_f = 8$. Both invariant manifolds cross $x$-axis for $x < -\mu$ (purple segment). Didymos and Dimorphos not to scale and represented as black triangle and dot, respectively. Dimorphos SOI is the dashed circle, $R_{\text{SOI}} = 183$ m.

(b) Structures associated to stable invariant manifold $\mathcal{W}^s_{L_1}$ of Fig. 3a within LD scalar field in the PCR3BP, field propagated from $f_0 = 0$ to $t_f = 8$. Bottom-left corner, stable invariant manifold $\mathcal{W}^s_{L_1}$ cut $\Gamma^s_{L_1}$ on $x\dot y$-plane as a red curve overlapped on the LD field. Top-right corner, level curves of gradient vector magnitudes of LD field.

**Figure 3:** LD scalar field features associated to stable invariant manifolds $\mathcal{W}^s_{L_1}$ of $L_1$ LO having Jacobi constant $C_j = 3.155\,086$.

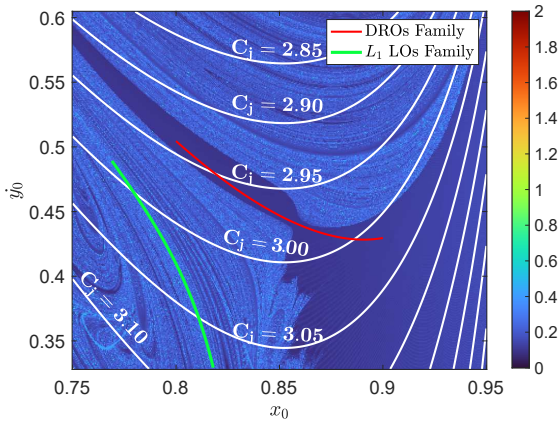
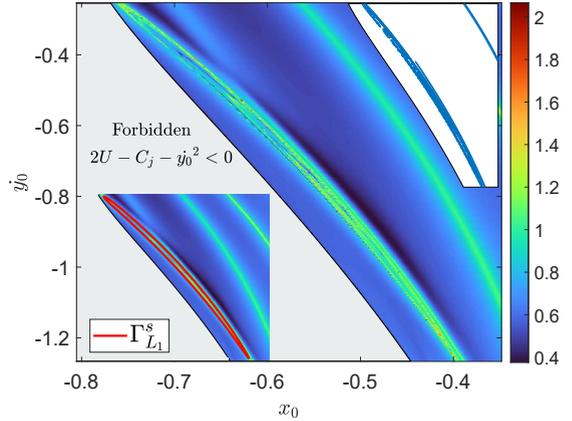

(a) Periodic orbit families overlapped to FTLE scalar field in the PCR3BP, Poincaré section $y_0 = \dot x_0 = 0$. Level curves of Jacobi constant $C_j$ are plotted in white. The smooth region visible in Fig. 1 and containing the DROs family is also detected by this variational diagnostic. The field exhibits rippled regions similar to those in Figs. 1 and 2.

(b) Structures associated to stable invariant manifold $\mathcal{W}^s_{L_1}$ of Fig. 3a within FTLE scalar field in the PCR3BP, field propagated from $f_0 = 0$ to $t_f = 8$. Bottom-left corner, stable invariant manifold $\mathcal{W}^s_{L_1}$ cut $\Gamma^s_{L_1}$ on $x\dot y$-plane as a red curve overlapped on the FTLE field. Top-right corner, level curves of gradient vector magnitudes of FTLE field.

**Figure 4:** FTLE scalar fields in the PCR3BP used as validation for results obtained with LD method.

the proximity of the Didymos double asteroid, hence requiring some form of trajectory control to temporarily remain there [10, 11].

Surprisingly, some additional spots featuring bounded motion are retained when the double asteroid is at the apocenter ($\theta_0 = \pi$) of its heliocentric orbit (see Fig. 6b). The first is a tiny portion of the channel seen in Figs. 1 and 5, which is still present in the top-left corner of the field. Sample p retraces the typical shape of these orbits. The second is the wider smooth area located on the right. Sample t displays the peculiar behavior of those solutions, which persistently revolve very close to Dimorphos, likely impacting with its surface multiple times. The other tested ICs (samples q, r, and s) immediately escape from Dimorphos. ICs of trajectories sampled in Fig. 6 are collected in Table 2.

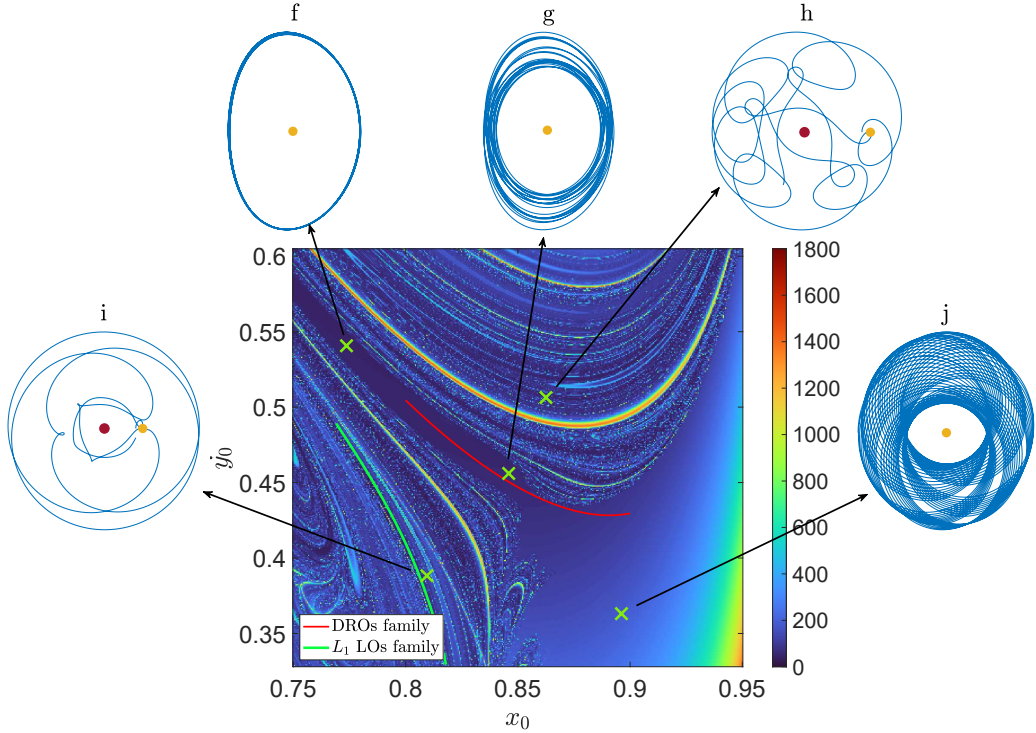

**Figure 5:** LD scalar field propagated in the PBER4BP with initial true anomaly $\theta_0 = 0$, Poincaré section $y_0 = \dot{x}_0 = 0$. DROs and $L_1$ LOs families in Fig. 1 reported as red and green curves, respectively. Samples f, g, and j evolve in bounded orbits. Samples h and i escape. Didymos and Dimorphos represented not to scale as red and yellow dots, respectively.

To better understand how SRP disrupts regions of bounded motion, a sequence of panels propagated in the PBER4BP when the Didymos system is close to perihelion ($\theta_0 = 0$) are proposed in Fig. 7. The panels are computed on grids of $100 \times 100$ points for increasing values of the parameter $\varepsilon = \{0.01, 0.1, 0.5, 1\}$ that multiplies coefficients $\alpha$ and $\beta$, so propagating the EoM with modified coefficients $\bar{\alpha} = \varepsilon\alpha$ and $\bar{\beta} = \varepsilon\beta$. The panels sequence clearly captures the transition from the results in Fig. 1 to those provided in Fig. 6a.

The dynamical origin of some selected features visible in Fig. 6a is investigated. Results are proposed in Fig. 8. Specifically, plots in Figs. 8a–8c show the trajectories of the purposely chosen orbits selected across abrupt changes of the LD scalar field. Exact ICs of such orbits are collected in Table 3. The difference by individual coordinates of the coupled orbits is shown in Fig. 8d. According to the results, the feature across samples 1 and 2 is not a separatrix but rather an artifact due to the presence of singularities in the EoM. In fact, the abrupt change is caused by a close encounter with D1 occurring at $f \approx 23$. Conversely, the abrupt changes featured in correspondence of orbits 3 and 4, and samples 5 and 6 are genuinely dynamical separatrices as clearly shown by the trajectory plots and the difference between coordinates.

Details about the correctness of the numerical integrations of the EoM are given in Fig. 9. Propagations are checked against a lower order integrator. Specifically,the Dormand–Prince 8th-order embedded Runge–Kutta (RK) method propagation scheme that is an adaptive step, 8th-order RK integrator with 7th-order error control [9]. Relative and absolute tolerances have been set to $10^{-12}$. In Fig. 9, the two panels on top provide the logarithm on base 10 of maximum errors in final position and velocity for Fig. 6a, while bottom plots show the same quantities but for Fig. 6b. Over the ten primaries revolutions time span considered, the two propagation schemes differ by errors smaller than $10^{-6}$ within regions exhibiting bounded motion. On the other hand, errors are larger at and close to the abrupt changes of the LD fields. In the bottom-right corner of the panels, both schemes struggle in providing reliable results. Furthermore,

Finding regions of bounded motion in binary asteroid environment using Lagrangian descriptors

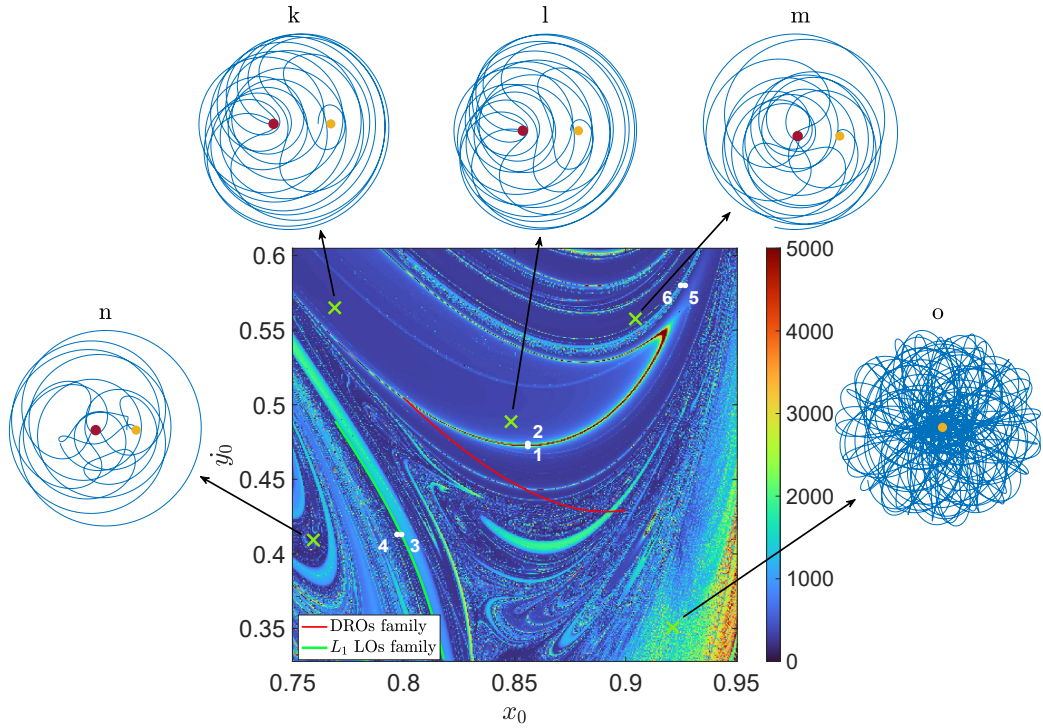

(a) Initial true anomaly $\theta_0 = 0$. Regions of bounded motion are almost lost. Sample o is the only one that does not escape.

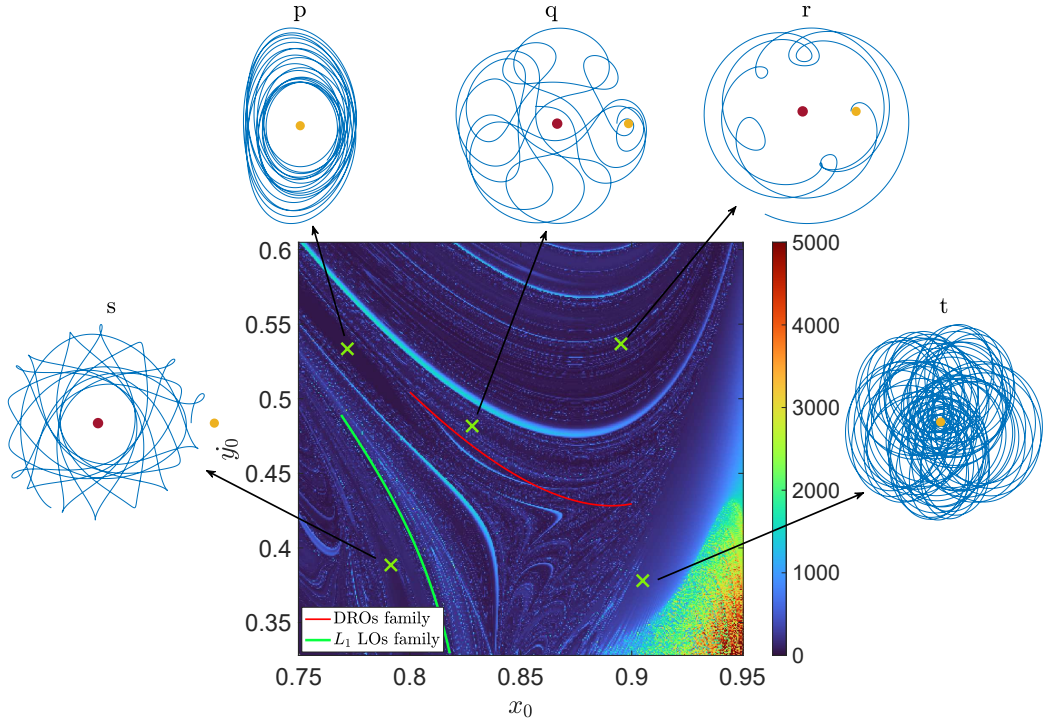

(b) Initial true anomaly $\theta_0 = \pi$. Samples p and t belong to different regions of bounded motion. Samples q, r, and s escape.

**Figure 6:** LD scalar field propagated in the PBER4BP perturbed by the SRP, Poincaré section $y_0 = \dot{x}_0 = 0$. DROs and $L_1$ LOs families in Fig. 1 reported as red and green curves, respectively. Didymos and Dimorphos represented not to scale as red and yellow dots, respectively.

**Table 2**
Initial conditions of sample orbits.

| Orbit | Initial condition at $f_0 = 0$ | | | | | Motion | Dynamics |
|---|---|---|---|---|---|---|---|
| | $x_0$ | $y_0$ | $\dot{x}_0$ | $\dot{y}_0$ | $\theta_0$ | | |
| a | 0.783 834 | 0 | 0 | 0.532 636 | - | Bounded | PCR3BP |
| b | 0.838 889 | 0 | 0 | 0.464 891 | - | Bounded | PCR3BP |
| c | 0.894 344 | 0 | 0 | 0.545 632 | - | Escape | PCR3BP |
| d | 0.765 415 | 0 | 0 | 0.403 230 | - | Escape | PCR3BP |
| e | 0.915 766 | 0 | 0 | 0.375 855 | - | Bounded | PCR3BP |
| f | 0.773 624 | 0 | 0 | 0.540 655 | 0 | Bounded | PBER4BP |
| g | 0.845 896 | 0 | 0 | 0.456 043 | 0 | Bounded | PBER4BP |
| h | 0.862 513 | 0 | 0 | 0.506 367 | 0 | Escape | PBER4BP |
| i | 0.809 460 | 0 | 0 | 0.388 298 | 0 | Escape | PBER4BP |
| j | 0.896 146 | 0 | 0 | 0.363 136 | 0 | Bounded | PBER4BP |
| k | 0.768 819 | 0 | 0 | 0.564 987 | 0 | Escape | PBER4BP with SRP |
| l | 0.848 298 | 0 | 0 | 0.488 671 | 0 | Escape | PBER4BP with SRP |
| m | 0.904 354 | 0 | 0 | 0.557 522 | 0 | Escape | PBER4BP with SRP |
| n | 0.759 209 | 0 | 0 | 0.409 313 | 0 | Escape | PBER4BP with SRP |
| o | 0.921 171 | 0 | 0 | 0.350 693 | 0 | Bounded | PBER4BP with SRP |
| p | 0.771 822 | 0 | 0 | 0.533 465 | $\pi$ | Bounded | PBER4BP with SRP |
| q | 0.828 078 | 0 | 0 | 0.481 758 | $\pi$ | Escape | PBER4BP with SRP |
| r | 0.895 145 | 0 | 0 | 0.536 783 | $\pi$ | Escape | PBER4BP with SRP |
| s | 0.791 441 | 0 | 0 | 0.388 575 | $\pi$ | Escape | PBER4BP with SRP |
| t | 0.904 955 | 0 | 0 | 0.378 067 | $\pi$ | Bounded | PBER4BP with SRP |

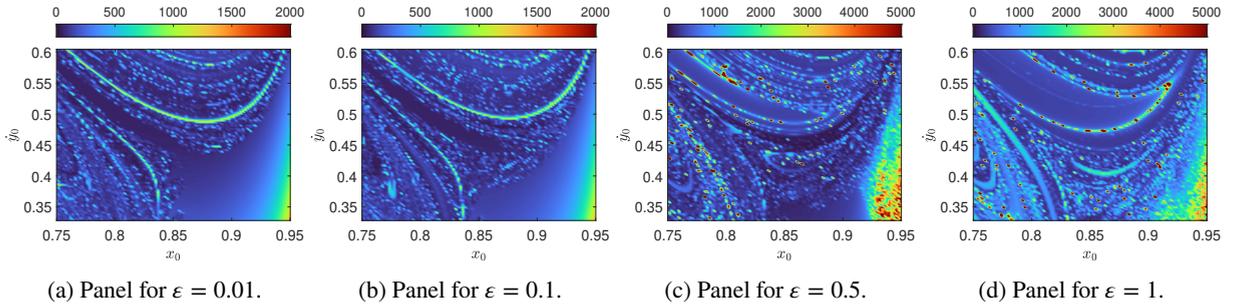

(a) Panel for $\varepsilon = 0.01$.    (b) Panel for $\varepsilon = 0.1$.    (c) Panel for $\varepsilon = 0.5$.    (d) Panel for $\varepsilon = 1$.

**Figure 7:** Sequence of panels of the LD scalar field propagated in the PBER4BP perturbed by increasing contributions of the four-body acceleration term and SRP, Poincaré section $y_0 = \dot{x}_0 = 0$. Panels computed on grids of $100 \times 100$ points.

**Table 3**
Initial conditions of orbits sampled across remarkable structures visible in Fig. 6a.

| Orbit | Initial condition at $f_0 = 0$ | | | | | Dynamics |
|---|---|---|---|---|---|---|
| | $x_0$ | $y_0$ | $\dot{x}_0$ | $\dot{y}_0$ | $\theta_0$ | |
| 1 | 0.856 000 | 0 | 0 | 0.472 000 | 0 | PBER4BP with SRP |
| 2 | 0.856 000 | 0 | 0 | 0.474 000 | 0 | PBER4BP with SRP |
| 3 | 0.799 000 | 0 | 0 | 0.413 000 | 0 | PBER4BP with SRP |
| 4 | 0.797 000 | 0 | 0 | 0.413 000 | 0 | PBER4BP with SRP |
| 5 | 0.927 000 | 0 | 0 | 0.580 000 | 0 | PBER4BP with SRP |
| 6 | 0.925 000 | 0 | 0 | 0.580 000 | 0 | PBER4BP with SRP |

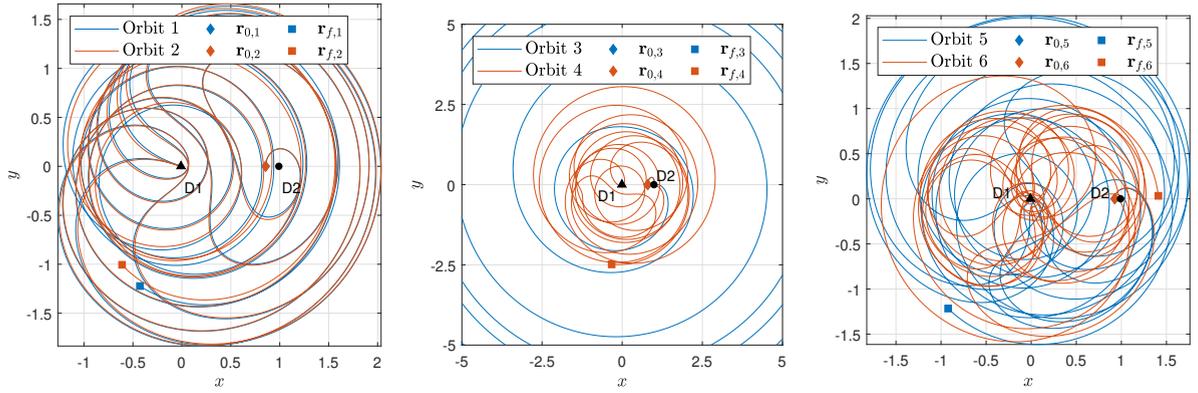

(a) Comparison between orbits 1 and 2. Orbits exhibit similar dynamical behavior, thus the correspondent feature in the field is a not a dynamical separatrix but rather an artifact due to the presence of singularities in the EoM.

(b) Comparison between orbits 3 and 4. Orbits exhibit different dynamical behavior, thus the correspondent feature in the field is a dynamical separatrix.

(c) Comparison between orbits 5 and 6. Orbits exhibit similar dynamical behavior, thus the correspondent feature in the field is a dynamical separatrix.

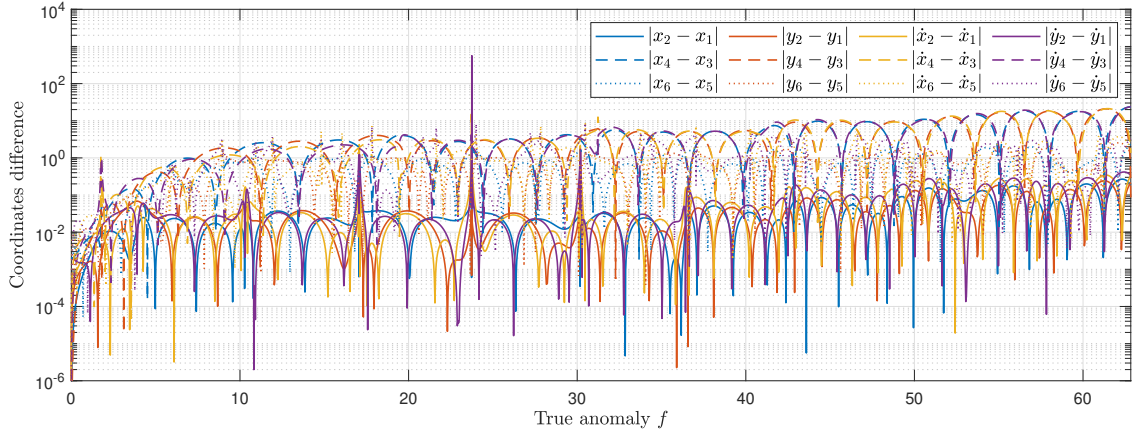

(d) Difference between coupled orbits sampled across remarkable structures of the LD field.

**Figure 8:** Investigation on dynamical origin of the LD field structures shown in Fig. 6a.

at least one of the two scheme cannot assure satisfaction of the tolerances set for some ICs. Such ICs are marked with white spots in Fig. 9. Overall, they are the 2.6% and 2.3% for Figs. 6a and 6b, respectively, of the 16 000 ICs propagated. They correspond to orbits performing close encounters with the primaries and are mostly concentrated in the bottom-right corner of the panels.

After a proper classification, the LD field effectively reveals structures that organize the phase space into regions of different dynamical behavior. This capability is retained when moving towards more representative, non-autonomous dynamical models that are poor of periodic solutions. Numerical experiments here not reported have shown that for longer integration intervals the LD field tends to reveal more information. Nonetheless, the arbitrary chosen final true anomaly $f_f = 20\pi$ is large enough to highlight most of the dynamical features. Particularly, the peculiar overwhelming strength of the SRP compared to the other gravitational pulls is well represented over the ten revolutions period. At the closest passage to the Sun, the SRP heavily modifies the LD field structure, but small regions of bounded motion could be potentially found for shorter integration intervals. LDs provide a way to explore the rich dynamics of non-autonomous models and proved to be an alternative to other chaos indicators adapted to astrodynamics for trajectory design purposes (e. g., FLI [14, 43], FTLE [15, 41], the variational theory for Lagrangian coherent structures [21, 32, 42], or the nonlinearity index [12, 25]).

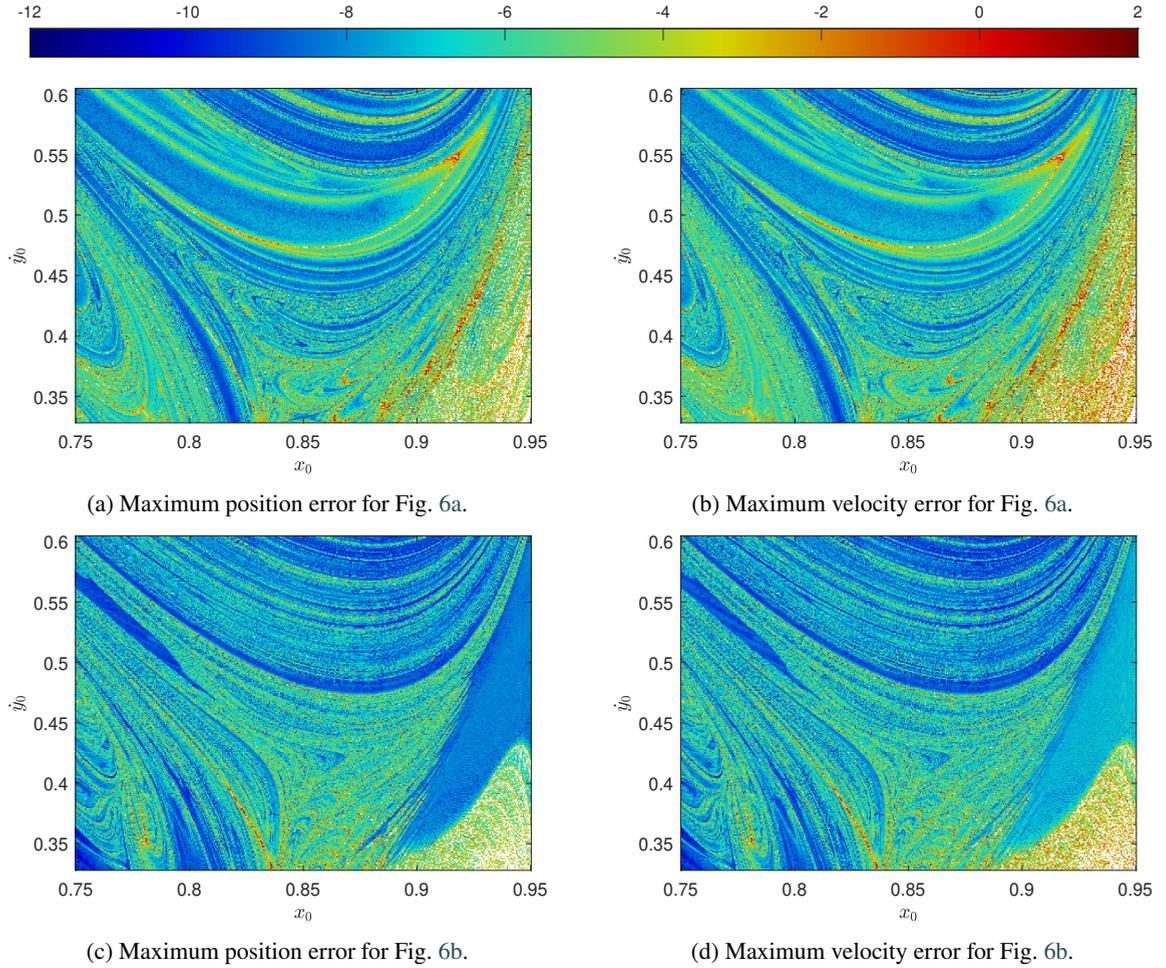

Figure 9: Correctness of numerical propagations. Check performed with an independent lower order integrator implementing DOPRI8 propagation scheme [9]. Logarithm on base 10 of maximum errors on final position and velocity of the same orbits presented in Fig. 6. White spots mark ICs for which at least one of the two propagation schemes could not satisfy the set tolerances.

## 5. Conclusion

This study looks into the effectiveness of Lagrangian descriptors in revealing phase space organizing structures in binary asteroid environments. They demonstrate success in finding regions of bounded motion in non-autonomous dynamical models of the (65803) Didymos system. Features highlighted by Lagrangian descriptor scalar fields are classified through a preliminary inspection in the planar circular restricted three-body problem. In the planar bi-elliptic restricted four-body problem, while bounded motion areas are readily captured in the absence of the solar radiation pressure, its contribution breaks down the majority of structures, so leading to a vast region of unbounded motion with rare exceptions. Ultimately, Lagrangian descriptors provide perceptive dynamical knowledge about the problem at hand and could be conveniently used in astrodynamics for preliminary trajectory design and quick exploration of the phase space. Nevertheless, a posteriori verification with variational methods is suggested due to Lagrangian descriptors incomplete reliability.

## CRediT authorship contribution statement

**Sebastiano Raffa:** Investigation, Methodology, Writing – Original draft preparation. **Gianmario Merisio:** Methodology, Writing – Original draft preparation. **Francesco Topputo:** Conceptualization of this study, Methodology, Writing – Draft revision, Funding acquisition.

## Declaration of competing interest

The authors declare that they have no known competing financial interests or personal relationships that could have appeared to influence the work reported in this paper.

## Acknowledgment

The authors would like to thank Dr. C. Giordano for the fruitful suggestions on how to compute the $\dot{\theta}$ derivative in Eq. (2). The authors would like to thank the anonymous reviewers for the helpful comments at the review stage. G.M. and F.T. would like to acknowledge the European Research Council (ERC) since part of this work has received funding from the ERC under the European Union's Horizon 2020 research and innovation programme (Grant Agreement No. 864697).

## References

[1] Benettin, G., Galgani, L., Strelcyn, J., 1976. Kolmogorov entropy and numerical experiments. Physical Review A 14, 2338. doi:10.1103/PhysRevA.14.2338.
[2] Broucke, R.A., 1968. Periodic orbits in the restricted three-body problem with Earth–Moon masses. Technical Report 32-1168. NASA, Jet Propulsion Laboratory, California Institute of Technology, Pasadena, California.
[3] Chen, H., Canalias, E., Hestroffer, D., Hou, X., 2020. Effective stability of quasi-satellite orbits in the spatial problem for Phobos exploration. Journal of Guidance, Control, and Dynamics 43, 2309–2320. doi:10.2514/1.g004911.
[4] Cheng, A.F., Atchison, J., Kantsiper, B., Rivkin, A.S., Stickle, A., Reed, C., Galvez, A., Carnelli, I., Michel, P., Ulamec, S., 2015. Asteroid impact and deflection assessment mission. Acta Astronautica 115, 262–269. doi:10.1016/j.actaastro.2015.05.021.
[5] Cheng, A.F., Rivkin, A.S., Michel, P., Atchison, J., Barnouin, O., Benner, L., Chabot, N.L., Ernst, C., Fahnestock, E.G., Kueppers, M., et al., 2018. AIDA DART asteroid deflection test: Planetary defense and science objectives. Planetary and Space Science 157, 104–115. doi:10.1016/j.pss.2018.02.015.
[6] Cincotta, P.M., Simó, C., 2000. Simple tools to study global dynamics in non-axisymmetric galactic potentials–i. Astronomy and Astrophysics Supplement Series 147, 205–228. doi:10.1051/aas:2000108.
[7] Curtis, H., 2020. Orbital Mechanics for Engineering Students: Revised Reprint. Butterworth-Heinemann. doi:10.1016/C2020-0-01873-6.
[8] Darriba, L.A., Maffione, N.P., Cincotta, P.M., Giordano, C.M., 2012. Comparative study of variational chaos indicators and ODEs' numerical integrators. International Journal of Bifurcation and Chaos 22, 1230033. doi:10.1142/S0218127412300339.
[9] Dormand, J.R., Prince, P.J., 1980. A family of embedded Runge–Kutta formulae. Journal of Computational and Applied Mathematics 6, 19–26. doi:10.1016/0771-050x(80)90013-3.
[10] Ferrari, F., Franzese, V., Pugliatti, M., Giordano, C., Topputo, F., 2021a. Preliminary mission profile of Hera's Milani CubeSat. Advances in Space Research 67, 2010–2029. doi:10.1016/j.asr.2020.12.034.
[11] Ferrari, F., Franzese, V., Pugliatti, M., Giordano, C., Topputo, F., 2021b. Trajectory options for Hera's Milani CubeSat around (65803) Didymos. The Journal of the Astronautical Sciences 68, 973–994. doi:10.1007/s40295-021-00282-z.
[12] Fossà, A., Armellin, R., Delande, E., Losacco, M., Sanfedino, F., 2022. Multifidelity orbit uncertainty propagation using Taylor polynomials, in: AIAA SCITECH 2022 Forum, pp. 0859, 1–16. doi:10.2514/6.2022-0859.
[13] Froeschlé, C., Lega, E., 1996. On the measure of the structure around the last kam torus before and after its break-up, in: Chaos in Gravitational N-Body Systems. Springer, pp. 21–31. doi:10.1007/978-94-009-0307-4_2.
[14] Froeschlé, C., Lega, E., Gonczi, R., 1997. Fast Lyapunov indicators. Application to asteroidal motion. Celestial Mechanics and Dynamical Astronomy 67, 41–62. doi:10.1023/A:1008276418601.
[15] Gawlik, E.S., Marsden, J.E., Du Toit, P.C., Campagnola, S., 2009. Lagrangian coherent structures in the planar elliptic restricted three-body problem. Celestial Mechanics and Dynamical Astronomy 103, 227–249. doi:10.1007/s10569-008-9180-3.
[16] Gil, P., Schwartz, J., 2010. Simulations of quasi-satellite orbits around Phobos. Journal of Guidance, Control, and Dynamics 33, 901–914. doi:10.2514/1.44434.
[17] Guzzo, M., Lega, E., 2015. A study of the past dynamics of comet 67P/Churyumov-Gerasimenko with fast Lyapunov indicators. Astronomy & Astrophysics 579, A79. doi:10.1051/0004-6361/201525878.
[18] Guzzo, M., Lega, E., 2018. Geometric chaos indicators and computations of the spherical hypertube manifolds of the spatial circular restricted three-body problem. Physica D: Nonlinear Phenomena 373, 38–58. doi:10.1016/j.physd.2018.02.003.
[19] Hadjighasem, A., Farazmand, M., Blazevski, D., Froyland, G., Haller, G., 2017. A critical comparison of lagrangian methods for coherent structure detection. Chaos: An Interdisciplinary Journal of Nonlinear Science doi:10.1063/1.4982720.
[20] Haller, G., 2001. Distinguished material surfaces and coherent structures in three-dimensional fluid flows. Physica D: Nonlinear Phenomena doi:10.1016/S0167-2789(00)00199-8.
[21] Haller, G., 2011. A variational theory of hyperbolic Lagrangian coherent structures. Physica D: Nonlinear Phenomena 240, 574–598. doi:10.1016/j.physd.2010.11.010.
[22] Haller, G., 2015. Lagrangian coherent structures. Annual Review of Fluid Mechanics doi:10.1146/annurev-fluid-010313-141322.
[23] Howell, K.C., 1984. Three-dimensional, periodic, 'halo' orbits. Celestial Mechanics 32, 53–71. doi:10.1007/bf01358403.
[24] Hyeraci, N., Topputo, F., 2010. Method to design ballistic capture in the elliptic restricted three-body problem. Journal of Guidance, Control, and Dynamics 33, 1814–1823. doi:10.2514/1.49263.

[25] Junkins, J.L., 1997. Von Karman lecture: Adventures on the interface of dynamics and control. Journal of Guidance, Control, and Dynamics 20, 1058–1071. doi:10.2514/2.4176.

[26] Laskar, J., 1993. Frequency analysis for multi-dimensional systems. Global dynamics and diffusion. Physica D: Nonlinear Phenomena 67, 257–281. doi:10.1016/0167-2789(93)90210-R.

[27] Laskar, J., 1999. Introduction to frequency map analysis, in: Hamiltonian systems with three or more degrees of freedom. Springer, pp. 134–150. doi:10.1007/978-94-011-4673-9_13.

[28] Lega, E., Froeschlé, C., 1997. Fast Lyapunov indicators comparison with other chaos indicators application to two and four dimensional maps, in: The Dynamical Behaviour of our Planetary System. Springer, pp. 257–275. doi:10.1007/978-94-011-5510-6_18.

[29] Lopesino, C., Balibrea-Iniesta, F., García-Garrido, V.J., Wiggins, S., Mancho, A.M., 2017. A theoretical framework for Lagrangian descriptors. International Journal of Bifurcation and Chaos 27, 1730001, 1–25. doi:10.1142/S0218127417300014.

[30] Luo, Z.-F., Topputo, F., Bernelli-Zazzera, F., Tang, G.-J., 2014. Constructing ballistic capture orbits in the real solar system model. Celestial Mechanics and Dynamical Astronomy 120, 433–450. doi:10.1007/s10569-014-9580-5.

[31] Mancho, A.M., Wiggins, S., Curbelo, J., Mendoza, C., 2013. Lagrangian descriptors: A method for revealing phase space structures of general time dependent dynamical systems. Communications in Nonlinear Science and Numerical Simulation 18, 3530–3557. doi:10.5194/npg-21-677-2014.

[32] Manzi, M., Topputo, F., 2021. A flow-informed strategy for ballistic capture orbit generation. Celestial Mechanics and Dynamical Astronomy 133, 1–16. doi:10.1007/s10569-021-10048-2.

[33] Michel, P., Kueppers, M., Sierks, H., Carnelli, I., Cheng, A.F., Mellab, K., Granvik, M., Kestilä, A., Kohout, T., Muinonen, K., et al., 2018. European component of the AIDA mission to a binary asteroid: Characterization and interpretation of the impact of the DART mission. Advances in Space Research 62, 2261–2272. doi:10.1016/j.asr.2017.12.020.

[34] Milani, A., Gronchi, G., 2010. Theory of orbit determination. Cambridge University Press. doi:10.1017/cbo9781139175371.002.

[35] Ming, X., Shijie, X., 2009. Exploration of distant retrograde orbits around Moon. Acta Astronautica 65, 853–860. doi:10.1016/j.actaastro.2009.03.026.

[36] Perozzi, E., Ceccaroni, M., Valsecchi, G.B., Rossi, A., 2017. Distant retrograde orbits and the asteroid hazard. The European Physical Journal Plus 132, 1–9. doi:10.1140/epjp/i2017-11644-0.

[37] Robin, I.A., Markellos, V.V., 1980. Numerical determination of three-dimensional periodic orbits generated from vertical self-resonant satellite orbits. Celestial Mechanics 21, 395–434. doi:10.1007/BF01231276.

[38] Ruiz-Herrera, A., 2015. Some examples related to the method of Lagrangian descriptors. Chaos: An Interdisciplinary Journal of Nonlinear Science doi:10.1063/1.4922182.

[39] Ruiz-Herrera, A., 2016. Performance of Lagrangian descriptors and their variants in incompressible flows. Chaos: An Interdisciplinary Journal of Nonlinear Science doi:10.1063/1.4966176.

[40] Russell, R., 2012. Global search for planar and three-dimensional periodic orbits near Europa. The Journal of the Astronautical Sciences 54, 199–226. doi:10.1007/BF03256483.

[41] Shadden, S.C., Lekien, F., Marsden, J.E., 2005. Definition and properties of Lagrangian coherent structures from finite-time Lyapunov exponents in two-dimensional aperiodic flows. Physica D: Nonlinear Phenomena 212, 271–304. doi:10.1016/j.physd.2005.10.007.

[42] Short, C.R., Blazevski, D., Howell, K.C., Haller, G., 2015. Stretching in phase space and applications in general nonautonomous multi-body problems. Celestial Mechanics and Dynamical Astronomy 122, 213–238. doi:10.1007/s10569-015-9617-4.

[43] Villac, B.F., 2008. Using FLI maps for preliminary spacecraft trajectory design in multi-body environments. Celestial Mechanics and Dynamical Astronomy 102, 29–48. doi:10.1007/s10569-008-9158-1.